# Real-Time Discrete SPAD Array Readout Architecture for Time of Flight PET

M.-A. Tétrault, *Student Member, IEEE*, É. Desaulniers Lamy, A. Boisvert, C. Thibaudeau,
F. Dubois, R. Fontaine, *Senior Member, IEEE,* J.-F. Pratte, *Member, IEEE*

*Abstract—* Single photon avalanche diode (SPAD) arrays have proven themselves as serious candidates for time of flight positron emission tomography (PET). Discrete readout schemes mitigate the low-noise requirements of analog schemes and offer very fine control over threshold levels and timing pickup strategies. A high optical fill factor is paramount to timing performance in such detectors, and consequently space is limited for closely integrated electronics. Nonetheless, a production, daily used PET scanner must minimize bandwidth usage, data volume, data analysis time and power consumption and therefore requires a real-time readout and data processing architecture as close to the detector as possible.

We propose a fully digital, embedded real-time readout architecture for SPAD-based detector. The readout circuit is located directly under the SPAD array instead of within or beside it to remove the fill factor versus circuit capabilities tradeoff. The overall real-time engine reduces transmitted data by a factor of 8 in standard operational mode. Combined with small local memory buffers, this significantly reduces overall acquisition dead time.

A prototype device featuring individual readout for 6 scintillator channels was fabricated. Timing readout is provided by a first photon discriminator and a 31 ps time to digital discriminator, while energy reading and event packaging is done using standard logic in real-time. The dedicated serial output line supports a sustained rate of 170k counts per second (CPS) in waveform mode, while the standard operational mode supports 2.2 MCPS.

## I. Introduction

TIME of flight is increasingly considered an essential path to improve contrast to noise ratio (CNR) in reconstructed positron emission tomography (PET) images [1, 2]. To do so, the data acquisition system (DAQ) must provide very sharp timing resolution.

Single photon avalanche diode (SPAD) devices have received great attention as a serious contender to achieve this goal. Like photomultiplier tubes, they have excellent gain and timing resolution but in addition are more compact, are immune to magnetic fields and require much lower voltage bias. Various SPAD array detector architectures are under investigation, the most common being the silicon photomultiplier (SiPM). These have reached PET coincidence resolutions below 200 ps in experimental setups with $LaBr_3$ [3] and LYSO [4] scintillators. However, SiPM devices require low-noise preamplifiers, their performance is limited by the noisiest SPAD in the array, and their dynamic range is fixed to a specific scale during acquisition.

Another approach reads SPAD in an array using digital electronics with array-wide first photon timestamp [5] or subgroup timestamps [6]; these introduce precise threshold control and allow very fine configuration of acquisition parameters. This, in turn, implies a tradeoff between photodetection fill factor and embedded circuitry complexity, affecting the effective photodetection efficiency. Furthermore, the missing external signal analysis steps require detailed data, increasing bandwidth usage and therefore latency, power consumption and acquisition dead time to varying degrees.

Whichever solution is selected, both analog and digital approaches require real-time device and system-level integration to minimize power consumption, dead time and physical size in the context of massive multi-channel environments such as a full PET system.

We propose a fully digital, real-time data acquisition (DAQ) architecture for a vertically stacked SPAD based PET detector. Unlike 2D electronic integration, vertical integration with through silicon via allows the presence of both a high fill factor SPAD array and a complex embedded system by placing the acquisition electronics under the photosensitive layer. The real-time embedded engine, placed on a dedicated layer, provides all the necessary modules for event detection and analysis, reducing the overall data throughput, power consumption and off-chip system complexity.

This paper begins with a short introduction to the overall detector specifications before focusing on the real-time DAQ. The detailed data flow is first presented, followed by the measurement methodology, and finally intrinsic experimental results from the key sub-components.

Manuscript submitted June 16th 2014. This work was supported in part by the Natural Sciences and Engineering Research Council of Canada (NSERC) and the Regroupement Stratégique en Microsystèmes du Québec (ReSMiQ).

M.-A. Tétrault, É. Desaulniers Lamy, A. Boisvert, F. Dubois, R. Fontaine and J.-F. Pratte are with the Department of Electrical and Computer Engineering, Université de Sherbrooke, Sherbrooke, QC, Canada

C. Thibaudeau is with the department of radiobiology and nuclear medicine, Université de Sherbrooke, Sherbrooke, QC, Canada

Contact information: e-mail: Marc-Andre.Tetrault@USherbrooke.ca

## II. Detector Architecture

To reach sub-millimeter spatial resolution, the overall detector uses $1.1 \times 1.1 \times 12$ mm$^3$ LYSO arrays with 0.1 mm gap, individually coupled to SPAD sub-matrices (Fig. 1, A). The SPAD devices used for this project are tiled with a 50 μm step, leading to an array of $22 \times 22$ per scintillator (Fig. 1, B), with a 100 μm gap between scintillator channels [7]. Individual quenching circuits with programmable hold-off and recharge delays are tiled under each SPAD (Fig. 1, D). The DAQ is located on its dedicated layer (Fig. 1, E). The first detector iteration supports 6 independent PET acquisition channels, for a total of 2904 SPAD units, with a 200 MHz system clock. A printed circuit board (PCB) supports the 3D electronics
(Fig. 1, F).

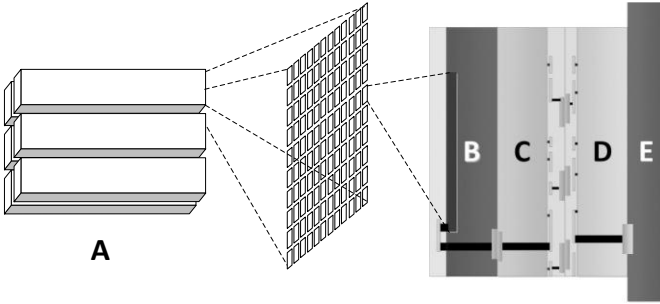

*Fig. 1. Detector module break down. A) Scintillation crystal array. B) 22 x 22 SPAD array implemented in Dalsa 0.8 μm HV CMOS. C) Quenching circuit layer, Global Foundries 130 nm CMOS. D) DAQ, Global Foundries 130 nm CMOS. E) PCB.*

## III. Real-time acquisition module

### A. Requirements

As for all PET DAQ systems, the DAQ needs to extract timestamp, energy and location information from incoming PET events. More importantly for time of flight measurements, the DAQ must eliminate or minimize its contribution to the overall timing jitter, independently of SPAD and quenching circuit contributions. For best PET singles detection performance, the DAQ must eliminate acquisition dead time as much as possible and quickly detect and discard dark counts that triggered false starts. Finally, to help mitigate non-linearities in the energy estimation [8], the DAQ should support counting multiple SPAD triggers during a PET event.

### B. Overview

To reach these design goals, the digital architecture is separated in 3 major blocks (Fig. 2): the timing pickup block, the DAQ, and the utilities module. Each PET channel has a dedicated timing and DAQ block, required by the 1:1 readout scheme. Post-sampling processing and the configuration interface are shared by every PET channel.

### C. Timing block

The timing pickup block is split into an OR-gate trigger tree, a delay-line first photon discriminator and a dual Vernier ring 20 ps TDC (Fig. 3). All module are asynchronous, except for the TDC handshake interface with the DAQ module.

The OR-gate trigger tree was designed using 4 stages of custom wired-OR cells. The tree is laid out by having the next stage gate in the geometric center of its input sources. This keeps wire connections as uniform as possible throughout the tree, reducing inter-SPAD skew, and therefore injected jitter.

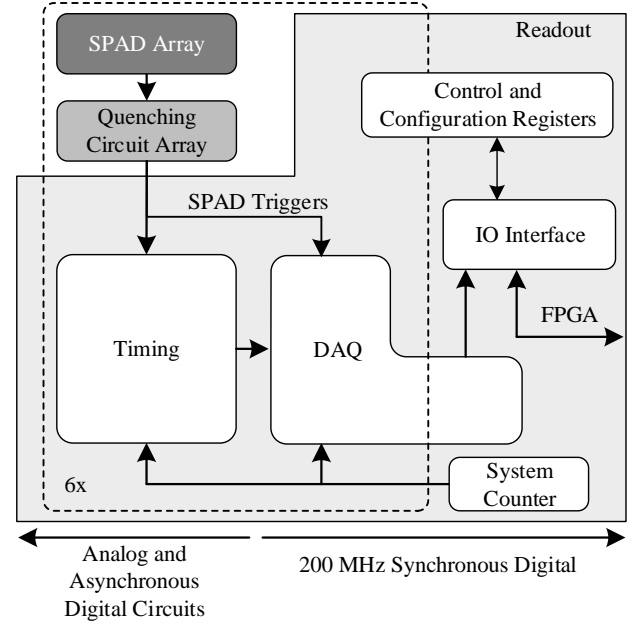

*Fig. 2. Architecture dataflow. The SPAD array, quenching circuit array and readout sections are layers C, D and E in figure 1, respectively. Dotted line represents channel physical separation.*

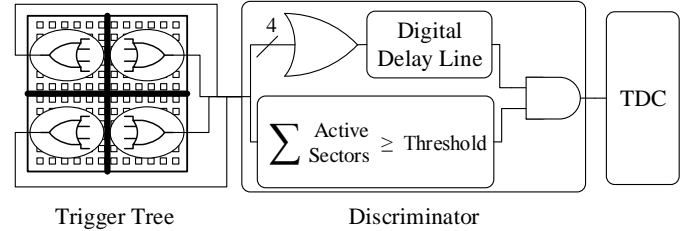

*Fig. 3. Timing block detail. On the left, an OR tree reduces triggers to a few signals. In the center, the delay line discriminator removes dark count induced false triggers. On the right, the TDC.*

A delay line discriminator is used to discard dark count false starts without affecting coincidence timing resolution [9]. At medium and high dark count rates, this is done at the cost of single PET event sensitivity, but with negligible to no cost at low dark count rates. In this implementation, the discriminator splits the SPAD array in 4 sectors to create the threshold comparison.

The dual Vernier ring TDC uses a 5 arbiter configuration with a 100 ns resolution time, shorter than 3 times the decay time of LYSO. It also features a reset port to quickly abort detections rejected by the DAQ unit (Fig. 4).

### D. DAQ dataflow

The DAQ block is fully synchronous logic, implementing all the necessary modules to monitor channels, detect events, and record PET events that fulfill trigger conditions (Fig. 4).

SPAD avalanche detection triggers cross into the synchronous domain firstly through a metastability protection stage and then through a single clocked pulse generation stage. The chain falls into self-reset for 1 clock period and then becomes immediately ready for the next trigger. Successive avalanches must occur at least 4 clock periods apart to be distinguished. The quenching circuit's combined hold-off and recharge time should thus be set to 4 clock periods or more. The SPAD hit map is then summed using 2 levels of parallel adders [10] to a 9-bit stream. The ADC-like sample stream is then written in a dual buffered, 64-word deep circular memory.

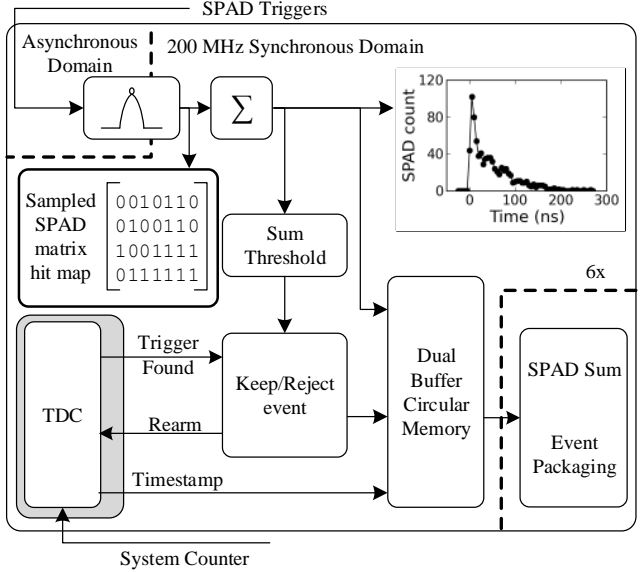

*Fig. 4. Acquisition dataflow. SPAD triggers pass into the synchronous domain through the metastability protection, and are them summed into a sample stream written to the circular memory. Accepted events are processed by the common integrator, packaged and then sent off chip for coincidence analysis.*

A simple level threshold acts as an additional discriminator concurrent to the timing circuit to identify in real time dark count triggers and low energy events. However, since the metastability protection and parallel sum introduce a 4 clock period latency with regard to the input triggers, the earliest abort signal to the channel is 4 clock periods.

Once an event is accepted, the control logic waits until 56 samples are written to the circular memory. The four following sample slots are overwritten with the TDC timestamp value and internal status flags, causing the only acquisition dead time in the system, and the dual buffer is switched. The four remaining memory slots thus provide the baseline prior to the signal's rising edge, for a total window of 60 samples or 300 ns.

Event data from all channels are funneled through the real-time parser which sums the total SPAD count. It then takes the final packaging decision: in regular event mode, only the address, timestamp, SPAD sum, and status flags are transmitted (Table I) to minimize latency and dead time. In waveform mode, the packager adds in the collected samples and data padding for off-chip transmission, but the format is about 8 times more bulky.

TABLE I
Event mode data fields description

| bits  | 63 32     | 31 22 | 21 16   | 15 0       |
|-------|-----------|-------|---------|------------|
| field | Timestamp | Flags | Address | SPAD Count |

### E. Utilities module

The utilities module regroups all other systems: various counters, the configuration and communication handshake module, the clock tree, the LVDS receivers (clock, commands and rough timer synchronization) and transmitters (data and command reply).

The data communication channels use a synchronous start/stop bit protocol to forgo a "chip select" line and minimize the package pin count. To accommodate different data types with various package sizes, the transmission engine uses a 16-bit data alignment and adds a 16-bit header containing data type and byte count information.

Finally, a low frequency internal trigger allows the user to simultaneously activate every quenching circuit on the die stack for functional tests and debug procedures.

## IV. MATERIALS AND METHODS

### A. Materials

The DAQ was fabricated with the Tezzaron/Global Foundries 3D mixed-signal 0.130 μm CMOS process. The prototype run included the quenching circuit tier and top-to-top bonding with the DAQ. This particular fabrication run has only the top tier (Fig. 1, D) thinned, so the circuit samples are wire-bonded to the PCB instead of soldered flip-chip (Fig. 5).

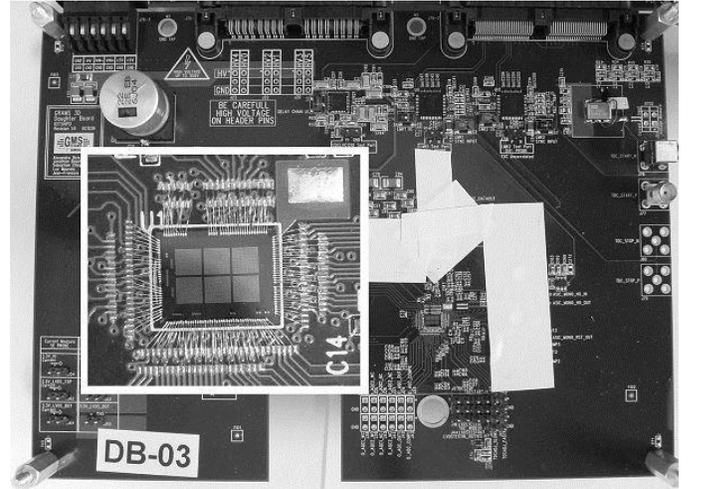

*Fig. 5. Sample die mounted on daughterboard PCB. The medallion zooms on the die, where the 6 input arrays are visible in the center.*

The test hardware includes a mainboard that houses an FPGA, external communications links, local memory, and power distribution. Die samples and all specific external test circuits are hosted on a daughterboard PCB (Fig. 5). A low jitter LVDS oscillator crystal provides the clock signal for the digital logic and communication and doubles up as the TDC stop signal.

Several tests also require an external start trigger. To this end, the daughterboard has two LMK01010 programmable clock buffer devices from Texas Instruments that act as trigger sources. One is for correlated triggers and uses the system clock. The other provides uncorrelated triggers from an independent oscillator crystal.

Finally, one of the acquisition channel's quenching circuit's input ports were all shorted together and bonded to the PCB. This simplifies experimental setups with the trigger sources and debug tools.

### B. Methods

#### 1) DAQ

The first series of tests verify architectural functionality compared to HDL functional simulations. These are all pass or fail tests: LVDS transceivers, command interface handshake, configuration bit assignments, digital threshold levels, and other debug sub-circuits. To confirm correct 3D interconnects, every quenching circuit is enabled in turn and pulsed by the slow internal trigger.

The sum tree feature is verified by enabling every quenching circuit in an acquisition channel and pulsing them with the slow internal trigger. The test is repeated with the correlated external trigger fixed to 25 MHz with 50% duty cycle. In this second test, the quenching circuit's hold-off and recharge analog delays are set in turn to the fastest and slowest values, approximately 20 ns and 1000 ns. In the fast case, because the rearming delay is shorter than the pulse rate, every quenching circuit is expected to trigger at each pulse. In the slow case, slight differences caused by process / voltage / temperature variations in the rearm analog circuitry will randomize occurrences over time, with peaks at the rising edge of the pulse.

#### 2) TDC

The TDC's DNL and INL were obtained with the statistical method. 400k triggers were harvested by using the external uncorrelated trigger and by activating a single quenching circuit in the DAQ channel. TDC behavior was further tested by observing a divided and delayed copy of the system clock. 5000 samples were acquired at 3 different delay settings: 800 ps, 1250 ps and 2750 ps.

#### 3) Trigger tree

The trigger tree skew and its resulting jitter were obtained with both simulation and experimental measurements. The simulated trigger tree propagation delays were obtained from a digital static timing analysis with parasitics for each input path. The physical delays were obtained by measuring each path's delay in turn with the internal TDC. In this case, the delay for each path is the mean from 400k triggers from the correlated trigger source. Skew refers here to the difference between the minimum and maximum values, while jitter is the root mean square of the delays. Simulated and measured data are centered on their respective means.

## V. RESULTS

### A. DAQ

The system runs successfully at the target 200 MHz frequency. The DAQ responds as expected to functionality tests, among them the LVDS transceivers, internal counters, and the configuration register access. Every quenching circuit individually responds to the internal trigger, confirming correct 3D interconnections between the DAQ tier and the quenching tier. The transmission engine endurance read and write tests correctly sends the data at 200 Mbps, confirming the 2.2 MCPS and 170 kCPS event rates in PET singles mode and waveform mode, respectively.

The count tree correctly accounts for all quenching circuit instances when using the internal trigger. With the external trigger, 15 quenching channels do not respond, likely due to failed bonding. Further rework was not done to avoid damaging the operational connections. At the shortest quenching setting, all 469 active quenching circuits consistently respond (Fig. 6). At the longest quenching setting, the slightly different rearm delays spread out the sum response in the expected irregular pattern, with triggers concentrated on the pulse's rising edge (Fig. 7).

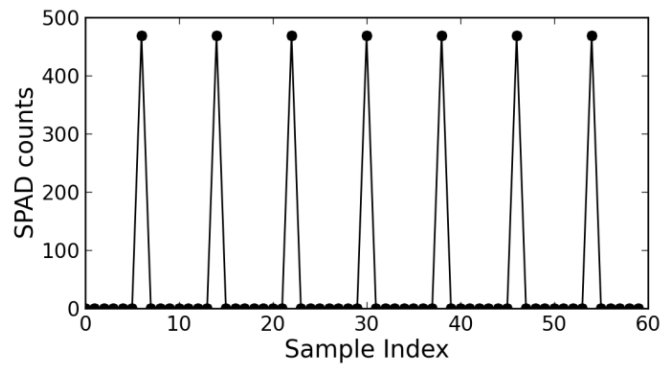

*Fig. 6. Waveform mode acquisition with the shortest quenching rearm setting and 25 MHz pulse input with 50% duty cycle. 469 quenching circuits respond at the pulse edge.*

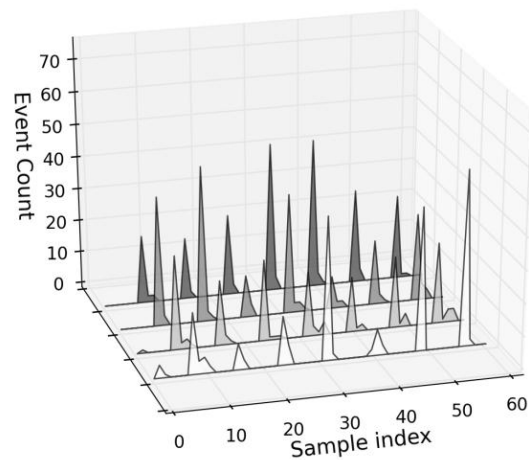

*Fig. 7. Four different waveform mode acquisitions with the longest quenching rearm setting and 25 MHz pulse input with 50% duty cycle. Trigger count is irregular due to the trigger being faster than the rearming delay, with an initial burst at the pulse edge.*

## B. TDC

The TDC's first design iteration achieves a 31 ps resolution, wider than the 20 ps design target. The maximum DNL is 2.15 LSB (Fig. 8) and the maximum INL is 9.6 LSB (Fig. 9). This performance was verified on 4 different die samples and will be revised in the next design iteration.

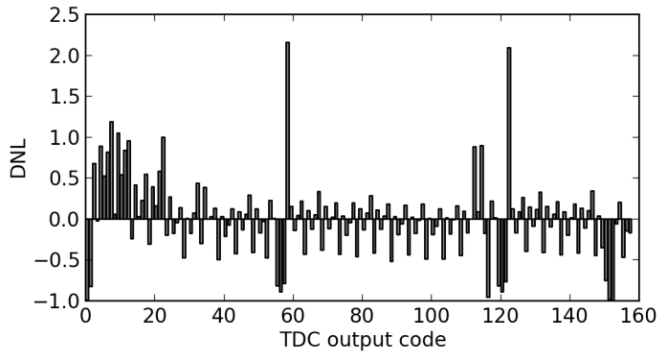

Fig. 8. DNL for the dual ring Vernier TDC

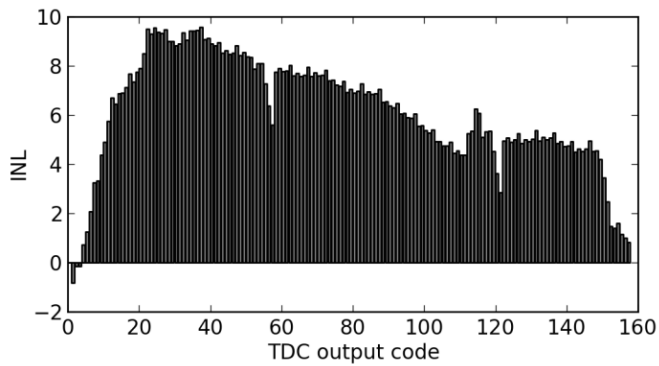

Fig. 9. Uncorrected INL for the dual ring Vernier TDC

The spike displacements in Fig. 10 further demonstrates the fabricated TDC's functionality at three different positions in the dynamic range.

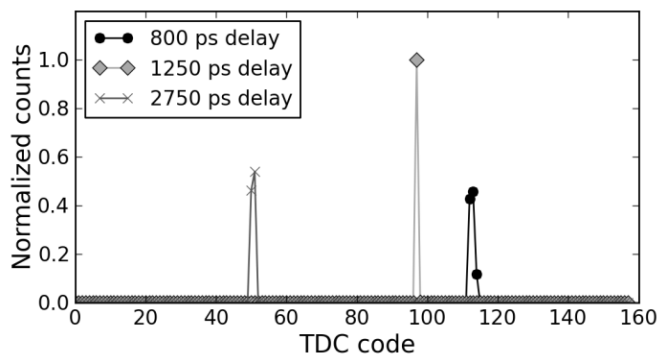

Fig. 10. TDC code distribution for 3 different delay values, with 5000 triggers in each acquisition.

## C. Trigger tree

Simulations predict 30 ps of skew in the 484 paths, producing 6 ps rms jitter in the trigger tree. Measurements report 72 ps of skew in the 469 acquired paths, minus 3 outliers, producing 18 ps rms jitter (Fig. 11). Although non-negligible, it is still smaller than the SPAD jitter [7] and smaller than 1 TDC bin. Skew-based jitter should improve by using matched length routing and shielded or coaxial traces in the integrated layout, at the cost of additional routing congestion.

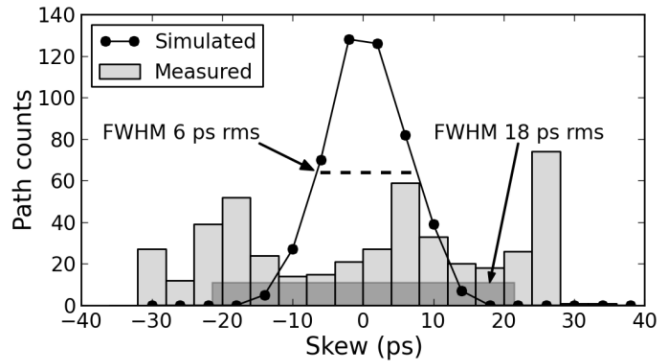

Fig. 11. Trigger tree skew histogram for the 484 possible paths between a SPAD input pulse and the TDC. Dotted line shows simulated FWHM, transparent grey box shows measured FWHM.

## VI. DISCUSSION

This paper focuses on intrinsic properties to isolate its contribution from other components in the system. For the proposed scintillator configuration, the overall timing resolution is expected in the order of 380 ps [9], and therefore the current readout jitter is acceptable. However, future research aims to reduce resolutions below 100 ps [3-4], which will require further improvements in the readout architecture and all other parts of the system.

The 1:1 scintillator readout should maximize spatial resolution since there is no decoding factor [11]. Finally, although the architecture supports successive SPAD triggers from individual units to mitigate non-linearities in energy measurement, its effectiveness will ultimately be limited by the hold-off and recharge time set by the user, in accordance to the SPAD array performance placed over the readout.

## VII. CONCLUSION

The readout architecture presented here takes advantage of the detector's 3D integration to include real-time data analysis and processing. This eliminates the need to move the data off-chip for feature extraction, significantly reducing dead time and transmission latency. The results present the intrinsic characterists of the architecture, and demonstrate that the readout itself maintains well the data quality. The fully integrated approach makes the overall detector ideal for high resolution, large axial field of view PET scanner, and the timing performance will not hinder TOF capability.

ACKNOWLEDGMENT

The authors would like to thank Moez Kanoun for his help with CAD tools and LVDS design, Luc Maurais and Jonathan Bouchard for the PCB board design, and Vincent-Philippe Rhéaume for help with the electronic setups.